\documentclass[a4paper,9pt]{article}
\usepackage{multicol}
\usepackage{authblk}
\usepackage[utf8]{inputenc}

\usepackage{url}
\usepackage{hyperref}
\usepackage{float}
\usepackage{amssymb}
\usepackage{amsmath}
\usepackage{array}
\usepackage{graphicx}

\newcommand{\apj}{Astrophysical Journal}

\usepackage{subfigure}
\graphicspath{{./images}}

\usepackage{placeins}

\begin{document}

\title{Bohmian singularity resolution and quantum relaxation in Bianchi type-I quantum cosmology}
 
\author{Vishal\thanks{vishal.phiitg@iitg.ac.in} }

\author{Malay K. Nandy\thanks{mknandy@iitg.ac.in (Corresponding Author)} }

\affil{Department of Physics, Indian Institute of Technology Guwahati, Assam, 781039, India} 
          
\date{(March 28, 2026)}            

\maketitle  
\begin{abstract}
We investigate cosmological singularity resolution and relaxation dynamics within the Bohmian mechanics via the plane-symmetric Bianchi type-I minisuperspace model in the Wheeler–DeWitt framework of quantum cosmology by constructing wavefunctions as Gaussian and Lorentzian wavepackets. Our analyses of the corresponding Bohmian trajectories reveal that Gaussian superposition predominantly yields classical singular solutions, with only a low fraction of small-amplitude cyclic trajectories. On the other hand, the Lorentzian wavepacket, characterized by the power-law momentum tail, generates stronger quantum potential barrier and a substantially rich velocity field, producing a significant fraction of non-singular bounce trajectories over extended volume ranges. We further examine quantum relaxation by evolving non-equilibrium distributions under the corresponding guidance dynamics. The Gaussian superposition exhibits laminar flow leading to boundary accumulation and incomplete relaxation, with non-monotonic decay of the $H$-function followed by saturation. In contrast, the Lorentzian wavepacket induces more complex trajectories, yielding monotonic decay of the $H$-function and better, though still incomplete, approach to Born-rule equilibrium. These results demonstrate that the inherent structure of the wave packet governs both singularity resolution and quantum relaxation through the nature of the Bohmian velocity field. \\


\end{abstract} 
\newpage
\tableofcontents

\section{Introduction}
A central challenge in developing a quantum theory of gravity is resolving the classical singularities—regions where general relativity predicts the breakdown of spacetime structure and known physical laws. The most prominent example is the big bang singularity, where the universe emerges from a state of vanishing volume and infinite energy density, with no meaningful extension of the spacetime before the big bang. This singularity was first identified within the Friedman–Lemaître–Robertson–Walker (FLRW) cosmological model, which assumes a perfectly homogeneous and isotropic universe~\cite{friedman1922krummung,lemaitre1927univers,robertson1935kinematics,walker1937milne}. Subsequent analysis of other models of the universe revealed that such singular behaviour is not an artifact of idealised symmetries but generic predictions of Einstein's theory under broad conditions~\cite{raychaudhuri1955relativistic}. Bianchi models provide the standard classification of anisotropic, spatially homogeneous spacetimes, relaxing the assumption of perfect isotropy~\cite{landau1971classical,ryan2015homogeneous,ellis2006bianchi}. These cosmological models serve as a more general testbed for gravitational dynamics and exhibit singular evolution, thereby confirming the existence of the singularity in classical gravity~\cite{collins1979singularities,ellis1999cosmological}.

At singularities, the classical description of spacetime ceases to be valid, suggesting the need for a more fundamental theory. This realization motivates the search for a quantum theory of gravity that can provide a consistent description of spacetime beyond the classical regime~\cite{wheeler1969superspace}. Among the various approaches to quantum gravity, one particularly well-studied direction is the canonical quantization of gravity, which was first introduced by DeWitt~\cite{dewitt1967quantum}. This quantization scheme aims to quantize the very geometry of spacetime by promoting the canonical variables of general relativity to quantum operators.

At the heart of quantum gravity lies the Wheeler–DeWitt (WDW) equation, $\hat{H} \Psi = 0$, which has long been regarded as the central equation of canonical quantum gravity. The solution of the WDW equation gives the wavefunction of the universe, which in principle can be used to study the quantum state of the universe~\cite{hartle1983wave}. 

The WDW equation has been widely applied in the study of cosmological singularities. Early investigations in minisuperspace models by DeWitt~\cite{dewitt1967quantum} suggested that wavefunctions vanishing at zero spatial volume could effectively avoid the classical big bang singularity. Subsequent works extended this approach by constructing specific wave packets and analyzing their implications for singularity avoidance in homogeneous cosmological models, including Bianchi universes and FLRW spacetimes~\cite{hartle1983wave,misner1969quantum,kiefer2019singularity,bouhmadi2014resolution}. These studies demonstrated that quantum effects within the Wheeler-DeWitt framework can replace classical singularities with scenarios such as oscillatory universes and bounces. Thus, the WDW equation provides a natural framework for investigating singularity resolution and testing whether quantum gravity yields a fundamentally non-singular description of the early universe.

However, the Wheeler-DeWitt (WDW) equation also presents conceptual challenges. Most notably, the wave function of the universe remains static, as the equation lacks an explicit time parameter. This leads to the well-known problem of time in quantum gravity~\cite{kuchavr2011time,isham1993canonical}. The problem originates from the fundamentally different roles time plays in general relativity (GR) and quantum mechanics (QM). In GR, time is dynamical, interwoven with space into a four-dimensional manifold whose geometry evolves according to the matter and energy content, admitting no globally preferred time coordinate. In QM, by contrast, time serves as an external, absolute parameter governing evolution of the system; it is not itself dynamical. Quantizing GR in the canonical formalism replaces the Schr\"{o}dinger equation with the timeless Wheeler-DeWitt (WDW) equation, in which the time coordinate disappears entirely. This timelessness in quantum gravity remains a central conceptual challenge and has motivated extensive investigation since the inception of quantum gravity~\cite{wald1980quantum,unruh1989time,gambini2009conditional,anderson2012problem,malkiewicz2020quantum,hohn2021trinity}.

The absence of time makes the interpretation of quantum cosmological dynamics particularly challenging and limits utility of the WDW framework for resolving singularities. In quantum cosmology, the problem acquires an additional layer: the universe as a closed system lacking an external observer, makes the problem of time unavoidable and directly links it to the description of cosmic evolution~\cite{rovelli1991time,page1983evolution}. This motivates alternative quantum formulations or interpretations that restore dynamical evolution within a consistent framework.

In this work, we focus on the singularity resolution and sidestep the longstanding problem of time in canonical quantum gravity by adopting the de Broglie--Bohm (dBB) pilot-wave interpretation~\cite{bohm1952suggested,bohm2006undivided,holland1995quantum,bohm1953proof}, a pragmatic approach employed in quantum cosmology. Unlike standard interpretations struggling with timeless wavefunctionals, dBB endows configuration space points with deterministic trajectories, $\dot{\mathbf{q}} = \nabla S/m$ guided by the phase $S=\Im[\ln\psi]$ of the wavefunction $\psi$. The defining feature is the quantum potential, $Q=-\frac{\nabla^2 R}{R}$ with $R=|\psi|$, absent from classical Hamilton--Jacobi theory, which produces quantum effects in the trajectories.

Bohmian cosmology has demonstrated significant success in singularity resolution. Pioneering studies of Gaussian wavepackets in isotropic FLRW minisuperspace revealed quantum bounce solutions~\cite{colistete2000gaussian,alvarenga2002quantum,pinto2013quantum,vicente2023bouncing}. These works identified possibilities of periodic universes, but the resulting bounces typically remain at Planck-scale volumes. Moreover, anisotropic quantum cosmological models have received limited attention within the Bohmian framework. Some studies have examined Bianchi type-I cosmologies, typically recovering bounce behaviors qualitatively similar to isotropic FLRW cases~\cite{pinto2000quantum,tavakoli2019bianchi}. 

Beyond providing a dynamical resolution to the timeless WDW equation, Bohmian mechanics offers another profound advantage: it treats the Born rule $\rho=|\psi|^2$ as an emergent equilibrium hypothesis rather than a fundamental postulate. Non-equilibrium distributions $\rho\neq|\psi|^2$ evolve according to the continuity equation, $\partial_t\rho + \nabla\cdot(\rho\mathbf{v})=0$ with velocities $\mathbf{v}=\nabla S/m$ fixed by the pilot wave. Quantum equilibrium, $\rho=|\psi|^2$  emerges dynamically through chaotic streamline mixing analogous to classical thermal relaxation. Antony Valentini pioneered the modern study of quantum relaxation, introducing the coarse-grained $H$-function
\begin{equation}
\bar{H}(t) = \int d\mathbf{q}\, \bar{\rho}\ln\left(\frac{\bar{\rho}}{|\psi|^2}\right)
\end{equation}
to quantify deviations from Born equilibrium~\cite{valentini1991signal1, valentini1991signal2}. Initial works examined quantum non-equilibrium using numerical simulations of simple systems, such as the two-dimensional infinite square well potential~\cite{valentini2005dynamical,towler2012time}, demonstrating relaxation to Born equilibrium $\rho=|\psi|^2$ via streamline mixing. In the same context, the two-dimensional harmonic oscillator was also studied, where the $H$-function first decays and then saturates~\cite{abraham2014long}. Later works on coupled harmonic oscillators showed that interactions and time-dependent coupling can retard or prevent complete relaxation for certain states~\cite{lustosa2021quantum,lustosa2023evolution}. 

Moreover, quantum nonequilibrium framework was applied to inflationary cosmology, demonstrating how incomplete relaxation during pre-inflationary phases can leave imprints on large-scale cosmic microwave background (CMB) anomalies~\cite{valentini2010inflationary,colin2013mechanism,colin2015primordial,colin2016robust,underwood2015quantum}. Recently, this investigation of quantum relaxation and validity of Born rule have been extended to high-energy collisions and black holes~\cite{valentini2025toward,kandhadai2020mechanism}. In quantum gravity contexts, Valentini argued that non-normalizable Wheeler-DeWitt solutions fundamentally lack Born-rule compliance at Planck scales, with relaxation emerging only during subsequent semiclassical expansion~\cite{valentini2023beyond}. The non-validity of Born rule in quantum gravity regime challenges the applicability of standard quantum mechanics at those scales.

This motivates our investigation of quantum relaxation alongside singularity resolution within the Wheeler-DeWitt framework of quantum gravity. Singularity resolution may couple intrinsically to equilibration through wavefunction superposition structure: persistent non-equilibrium ($\bar{H}(t)>0$) from laminar streamline flows (simple wavepackets) permits trajectories reaching the classical singularity, while chaotic superpositions driving relaxation ($\bar{H}(t)\to0$) also generate quantum potential barriers $Q=-\nabla^2R/R$ which repel trajectories from collapse regions. In this paper, we test this mechanism using Gaussian and Lorentzian wavepackets in the Bianchi type-I minisuperspace model. 

For the Gaussian and Lorentzian superpositions, we compute Bohmian trajectories and $H$-function to study singularity resolution and quntum non-equilibrium relaxation. Our central results demonstrate that Lorentzian wavepackets---with characteristic power-law tails supporting higher-$k$ modes---produce substantially superior singularity resolution compared to Gaussian envelopes, where most trajectories remain singular. Relaxation analysis reveals incomplete decay of the $H$-function in both cases, suggesting resistance to full equilibration in quantum gravity. For Lorentzian wave packets, the higher flow complexity giving singularity resolution correlates well with improved statistical mixing, leading to less departure from equilibration with respect to the Gaussian case.

The rest of the paper is organised as follows. Section~\ref{sec2} shows the WDW quantization of plane symmetric Bianchi type-I universe and the structure of Bohmian cosmology for minisuperspace models. In Sections~\ref{sec3} and~\ref{sec3a}, we compute Bohmian trajectories for Gaussian and Lorentzian superpositions, respectively. Section~\ref{sec4} gives an introduction about non-equilibrium quantum mechanics in cosmology followed by relaxation dynamics for both types of wavepackets. In Section~\ref{sec5}, we discuss our results and conclusions.
\section{Plane-symmetric background and de Broglie-Bohm interpretation}
\label{sec2}
Bianchi universes~\cite{ellis1999cosmological,ryan2015homogeneous} are the simplest classical anisotropic models which are homogenous but do not respect spatial isotropy. A particularly interesting subclass of Bianchi Type-I models is the plane symmetric or ellipsoidal universe, where two of the three spatial directions expand (or contract) identically, while the third evolves independently. The line element for the plane symmetric Bianchi Type-I universe is given by
\begin{equation}
ds^2 = - N dt^2  +a^2(t) (dx^2 +  dy^2) + b^2(t) dz^2,
\label{metric}
\end{equation}
where $a(t)$ and $b(t)$ are scale factors along the three different axes.

We analyze this anisotropic Bianchi type-I model within the minisuperspace Wheeler-DeWitt framework~\cite{WheelerQG,dewitt1967quantum}, focusing on early-universe dynamics. The determinant of the metric is $\sqrt{h} = a^2 b$, with extrinsic curvature scalar $K = -\frac{1}{N}\left(2\frac{\dot{a}}{a} + \frac{\dot{b}}{b}\right)$ and spatial Ricci scalar ${}^3R = 0$. We introduce logarithmic coordinates,
$\alpha = \log(a^2 b)$ and $\beta = \frac{1}{2}\log\left(\frac{b^2}{a^2}\right)$, where $\alpha$ represents the volume and $\beta$ parametrizes the anisotropy ($\beta=0$ recovers isotropy). With these variables, the action takes the form
\begin{equation}
S = \frac{1}{4 \pi G} \int dt \, d^3x  \frac{e^{3 \alpha}}{N}  \left( \dot{\beta}^2-\dot{\alpha}^2 \right),
\end{equation}
and the corresponding Hamiltoniam reads
\begin{equation}
H = p^{2}_{\beta} - p^{2}_{\alpha},
\end{equation}
where $p_\alpha$ and $p_\beta$ are momenta conjugate to the variables $\alpha$ and $\beta$, respectively.

We now quantize this minisuperspace model canonically by promoting the classical momenta to quantum operators, $p_{\beta} \rightarrow -i\hbar\frac{\partial}{\partial \beta},\quad p_{\alpha} \rightarrow -i\hbar\frac{\partial}{\partial \alpha}$, and by substituting into the Hamiltonian constraint $H=0$. This yields the Wheeler-DeWitt equation,
\begin{equation}
\left[
\frac{\partial^2}{\partial \alpha^2}
-\frac{\partial^2}{\partial \beta^2}
\right] \Psi(\alpha, \beta) = 0.
\label{wdeq}
\end{equation}

In order to obtain the solution of the equation we employ separation of variables and write the wavefunction as
\begin{equation}
\Psi(\alpha, \beta) = \int_{-\infty}^{\infty} F(k)\, A_k(\alpha)\, B_k(\beta)\, dk,
\end{equation}
where $F(k)$ is the  mode distribution function of separation constant $k$. The superposition of left and right moving components of volume and anisotropy are written as
\begin{eqnarray}
A_k(\alpha) = a_1 e^{i k \alpha} + a_2 e^{-i k \alpha}\\
B_k(\beta) = b_1 e^{i k \beta} + b_2 e^{-i k \beta}.
\end{eqnarray}
We consider the wave function
\begin{equation}
\Psi(\alpha,\beta) = \int F(k) \, e^{i k \beta} \left( e^{i k \alpha} + e^{-i k \alpha} \right) dk,
\label{wffk}
\end{equation}
featuring a superposition in the volume coordinate $\alpha$. This structure encodes a quantum combination of expanding $e^{i k \alpha}$ and contracting $e^{-i k \alpha}$ universe branches, while $e^{i k \beta}$ governs anisotropic evolution.

A quantum theory that can be consistently implemented in the quantum cosmology scenario is the de Broglie-Bohm (dBB) quantum theory~\cite{bohm1952suggested,bohm2006undivided,holland1995quantum}.
In the case of homogeneous minisuperspace models—characterized by a finite number of degrees of freedom—the general form of the associated Wheeler–DeWitt (WDW) equation is~\cite{colistete1998singularities,colistete2000gaussian}
\begin{equation}
- \frac{1}{2} f_{\rho \sigma}(q_\mu) \frac{\partial^2 \Psi(q)}{\partial q_\rho \partial q_\sigma} + U(q_\mu) \Psi(q) = 0,
\label{wdb}
\end{equation}
where $f_{\rho \sigma}(q_\mu)$ is the DeWitt supermetric of the
model, whose inverse is denoted by $f^{\rho \sigma}(q^\mu)$. By writing
the wave function in its polar form $\Psi = R \exp (iS/\hbar)$  , the complex equation~\ref{wdb} decouples into two real equations:
\begin{equation}
\frac{1}{2} f_{\rho \sigma}(q_\mu) \frac{\partial S}{\partial q_\rho} \frac{\partial S}{\partial q_\sigma} + U(q_\mu) + Q(q_\mu) = 0,
\label{hj}
\end{equation}

\begin{equation}
f_{\rho \sigma}(q_\mu) \frac{\partial}{\partial q_\rho} \left( R^2 \frac{\partial S}{\partial q_\sigma} \right) = 0,
\label{ct}
\end{equation}
where 
\begin{equation}
Q(q_\mu) = - \frac{1}{2R} f^{\rho \sigma} \frac{\partial^2 R}{\partial q_\rho \partial q_\sigma}
\label{qp}
\end{equation}
is called the quantum potential. equation~\ref{hj} is a Hamilton-Jacobi type equation for a particle submitted to an external potential which is the classical potential plus a new quantum potential $Q(q_\mu)$ responsible for the quantum effects. equation~\ref{ct} is a continuity equation which admits the interpretation of probability conservation if $R^2$ is identified with the probability density. In the dBB interpretation, the motion of the system’s configuration variables $q^{\mu}$ is governed by the guidance relations,
\begin{equation}
\frac{\partial S}{\partial q_\rho} = p^\rho = f_{\rho \sigma} \dot{q}^\sigma,
\label{gd}
\end{equation}
which connect the conjugate momenta directly to the phase $S$ of the wavefunction. The phase can be obtained for wavefunciton $\Psi$ using
\begin{equation}
\frac{d S}{d q_\rho} =  \frac{i}{2 \, \Psi^{*}\Psi} \left(\Psi \frac{\partial \Psi^{*}}{\partial q_\rho}-\Psi^{*}\frac{\partial \Psi}{\partial q_\rho}\right).
\label{phase}
\end{equation}
In the next section, we employ these equations to compute Bohmian trajectories for our cosmological model.

\section{Gaussian superposition and Bohmian trajectories}
\label{sec3}
We now construct a Gaussian superposition of the plane-wave solution, with $F(k) = e^{-\frac{(k-k_0)^2}{\sigma^2}}$ in equation~\ref{wffk}, and analyze the dynamics through the de Broglie--Bohm interpretation of quantum mechanics. The wavefunction for this superposition reads
\begin{equation}
\Psi_G(\alpha, \beta) = \sigma \sqrt{\pi}
\left[
    e^{-\frac{\sigma^2}{4}(\alpha + \beta)^2} e^{i k_0 (\alpha + \beta)}
    +
    e^{-\frac{\sigma^2}{4}(\alpha - \beta)^2} e^{-i k_0 (\alpha - \beta)}
\right].
\label{gaussian-wavefunction}
\end{equation}
Using equation~\ref{gd}, the guidance relations for $\alpha$ and $\beta$ turn out to be
\begin{equation}
\frac{d S}{d \beta} = p_\beta = 8 \frac{d \beta}{d t}
\label{gbeta}
\end{equation}
and
\begin{equation}
\frac{d S}{d \alpha} = p_\alpha = -8 \frac{d \alpha}{d t}.
\label{galpha}
\end{equation}
Calculating the phase derivatives using equation~\ref{phase} and plugging into the above guidance relations, two differential equations are obtained for the evolution of $\alpha$ and $\beta$, given by
\begin{equation}
\frac{d \alpha}{d t} = \frac{1}{16} \left[
\frac{
\sigma^2 \beta \sin\left(2  k_0 \alpha \right) + 2 k_0 \sinh\left(\sigma^2 \alpha \beta \right)
}{
\cosh\left(\sigma^2 \alpha  \beta \right) + \cos\left(2 k_0 \alpha \right)
} \right]
\label{dalphadt}
\end{equation}
and
\begin{equation}
\frac{d \beta}{dt} = \frac{1}{16} \left[
\frac{
-\sigma^2 \alpha  \sin\left(2 k_0 \alpha \right) + 2 k_0
\cosh\left(\sigma^2 \alpha \beta\right) + 2 k_0  \cos\left(2 k_0 \alpha \right)
}{
\cosh\left(\sigma^2 \alpha  \beta \right) + \cos\left(2 k_0 \alpha \right)
}\right].
\label{dbetadt}
\end{equation}

\begin{figure}[ht!]
    \begin{center}
    {\includegraphics[width=0.75 \textwidth,height =7 cm]{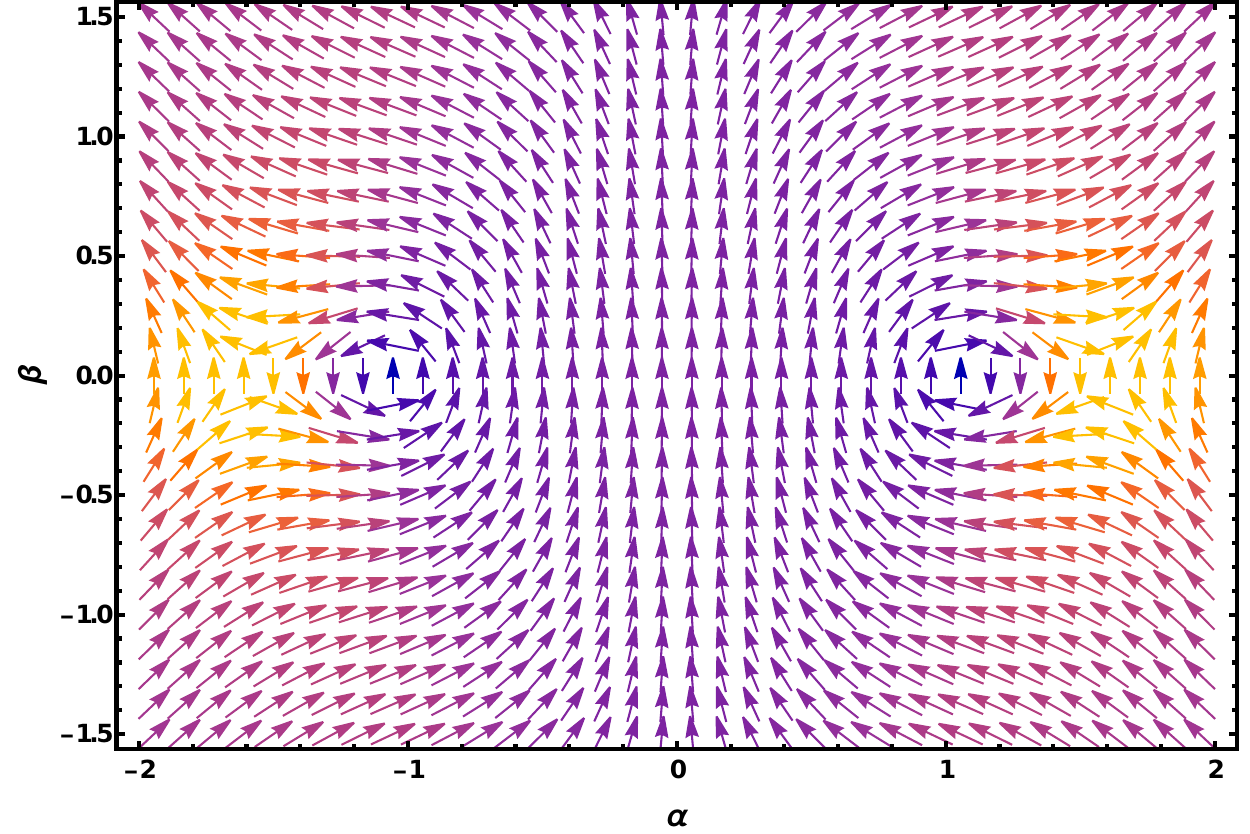}}\hfill%
     \end{center}
     \caption{Field plot of the guidance equations~\ref{dalphadt},~\ref{dbetadt} showing Bohmian trajectories for the Gaussian superposition case with $\sigma = k_0 = 1$. In the left sector ($\alpha < 0$), most trajectories converge toward the past singularity, while in the right sector ($\alpha > 0$), trajectories converge toward the future singularity. The color gradient indicates regions of rapid velocity flow. A few small circular loops near the isotropy region ($\beta \approx 0$) represent oscillating universes in both halves of the configuration space.}
    \label{fig:gaussian-vplot}
\end{figure}

Fig.~\ref{fig:gaussian-vplot} shows the field plot of the Bohmian trajectories for the above guidance equations~\ref{dalphadt},~\ref{dbetadt}. The line $\alpha = 0$ serves as a critical divider in configuration space, separating left-moving (contracting) and right-moving (expanding) sectors corresponding to the plane wave solutions of the Wheeler-DeWitt equation.

\FloatBarrier
In the left half of the configuration space ($\alpha < 0$), numerous trajectories initiate from classical singularities characterized by arbitrarily small volume ($\alpha \to -\infty$) and large anisotropy ($\beta < 0$, corresponding to $b^2 < a^2$). These universes initially expand classically and reache an isotropic phase, but due to quantum effects, the trajectories bend towards regions of small volume and large anisotropy ($\beta > 0$, or $b^2 > a^2$). 

Some of the trajectories show oscillating ``bounce'' universes that completely avoid the singularity. These closed orbits remain confined to finite volume regions with small anisotropies, undergoing periodic expansion-contraction cycles. The spatial scales of these cyclic universes remain extremely small, much below the observable universe's current size. 

In the right half ($\alpha > 0$), there are some oscillating universes symmetric to the left half of the configuration space, but most of the trajectories originate from arbitrarily large volumes ($\alpha \to +\infty$) with high anisotropy ($\beta < 0$), undergoing collapse towards near-isotropic configurations, and subsequently expand towards the future singularity. 

Thus, for the Gaussian superposition, we found that most of the trajectories exhibit classical behaviour and quantum effects are prominent in very few small regions. 

As the trajectories in the Bohmian approach are guided by the wavefunction, which is determined by the superposition characterised by the Gaussian function $F(k)$, we shall construct a wavepacket with a Lorentzian superposition in the next section and check how it changes the nature of the trajectories.

\section{Lorentzian Superposition and Bohmian trajectories}
\label{sec3a}
We now consider a Lorentzian wavepacket with the superposition function $F(k) = \frac{\gamma}{\pi[(k-k_0)^2 + \gamma^2]}$, in order to probe the high-momentum behavior near the classical singularity in the minisuperspace~\ref{metric}. Unlike Gaussian profiles that exponentially suppress large-$k$ modes, this $1/k^2$ power-law decay permits accounting for significantly high-momentum fluctuations essential for robust singularity resolution. Using the Fourier transform of Lorentzian distribution, 
\[
\int_{-\infty}^{\infty} 
e^{ikx}\,
\frac{\gamma}{\pi\left((k-k_0)^2+\gamma^2\right)}
\, dk
=
e^{i k_0 x - \gamma |x|},
\]
the wavefunction from equation~\ref{wffk} takes the form
\begin{equation}
\Psi(\alpha, \beta) = \frac{\gamma}{\pi}
\left[
     e^{i k_0 (\alpha + \beta)-\gamma |(\alpha + \beta)|}
    +
    e^{-i k_0 (\alpha - \beta)-\gamma |(\alpha - \beta)|}
\right],
\label{lorentzian-wavefunction}
\end{equation}
which accounts for the superposition of left and right moving plane waves in the variables $\alpha$ and $\beta$. The parameter $k_0$ determines the oscillatory phase of the state and $\gamma$ governs the localization scale. Moreover, presence of the absolute value $|\alpha\pm\beta|$ introduces abrupt change in the derivatives of the wavefunction, which can lead to sharp changes in the phase gradient along lines $\alpha+\beta=0$ and $\alpha-\beta=0$. Consequently, this influences the structure of the Bohmian velocity field and the associated trajectories. Using the derivatives
\begin{equation}
\frac{\partial \Psi} {\partial \alpha} =
\frac{\gamma}{\pi}
\left[
e^{i k_0 (\alpha+\beta)-\gamma|\alpha+\beta|}
\left(i k_0 -\gamma\,\mathrm{Sign}(\alpha+\beta)\right)
+
e^{-i k_0 (\alpha-\beta)-\gamma|\alpha-\beta|}
\left(-i k_0 -\gamma\,\mathrm{Sign}(\alpha-\beta)\right)
\right]
\end{equation}
and 
\begin{equation}
\frac{\partial \Psi}{\partial \beta} =
\frac{\gamma}{\pi}
\left[
e^{i k_0 (\alpha+\beta)-\gamma|\alpha+\beta|}
\left(i k_0 -\gamma\,\mathrm{Sign}(\alpha+\beta)\right)
+
e^{-i k_0 (\alpha-\beta)-\gamma|\alpha-\beta|}
\left(i k_0 +\gamma\,\mathrm{Sign}(\alpha-\beta)\right)
\right],
\end{equation}
the Bohmian guidance equations for the configuration variables $\alpha$ and $\beta$ turn out to be
\begin{equation}
\frac{d \alpha}{dt} = \frac{1}{16} \left[\frac{\left( \mathrm{Sign}(\alpha + \beta) - \mathrm{Sign}(\alpha - \beta) \right) \gamma \sin(2 k_0 \alpha) + 2 k_0 \sinh\left(\gamma \left( | \alpha + \beta | - | \alpha - \beta | \right) \right)}{\cosh\left(\gamma \left( | \alpha + \beta | - | \alpha - \beta | \right)\right) + \cos(2 k_0 \alpha)}\right]
\label{dadtL}
\end{equation}
and
\begin{equation}
\frac{d \beta}{dt} = \frac{1}{16} \left[ \frac{-\left( \mathrm{Sign}(\alpha + \beta) + \mathrm{Sign}(\alpha - \beta) \right) \gamma \sin(2 k_0 \alpha) + 2 k_0 \cosh\left(\gamma \left( | \alpha + \beta | - | \alpha - \beta | \right) \right) + 2 k_0 \cos(2 k_0 \alpha)}{\cosh\left(\gamma \left( | \alpha + \beta | - | \alpha - \beta | \right)\right) + \cos(2 k_0 \alpha)} \right].
\label{dbdtL}
\end{equation}

\begin{figure}[ht!]
    \begin{center}
    {\includegraphics[width=0.75 \textwidth,height =7 cm]{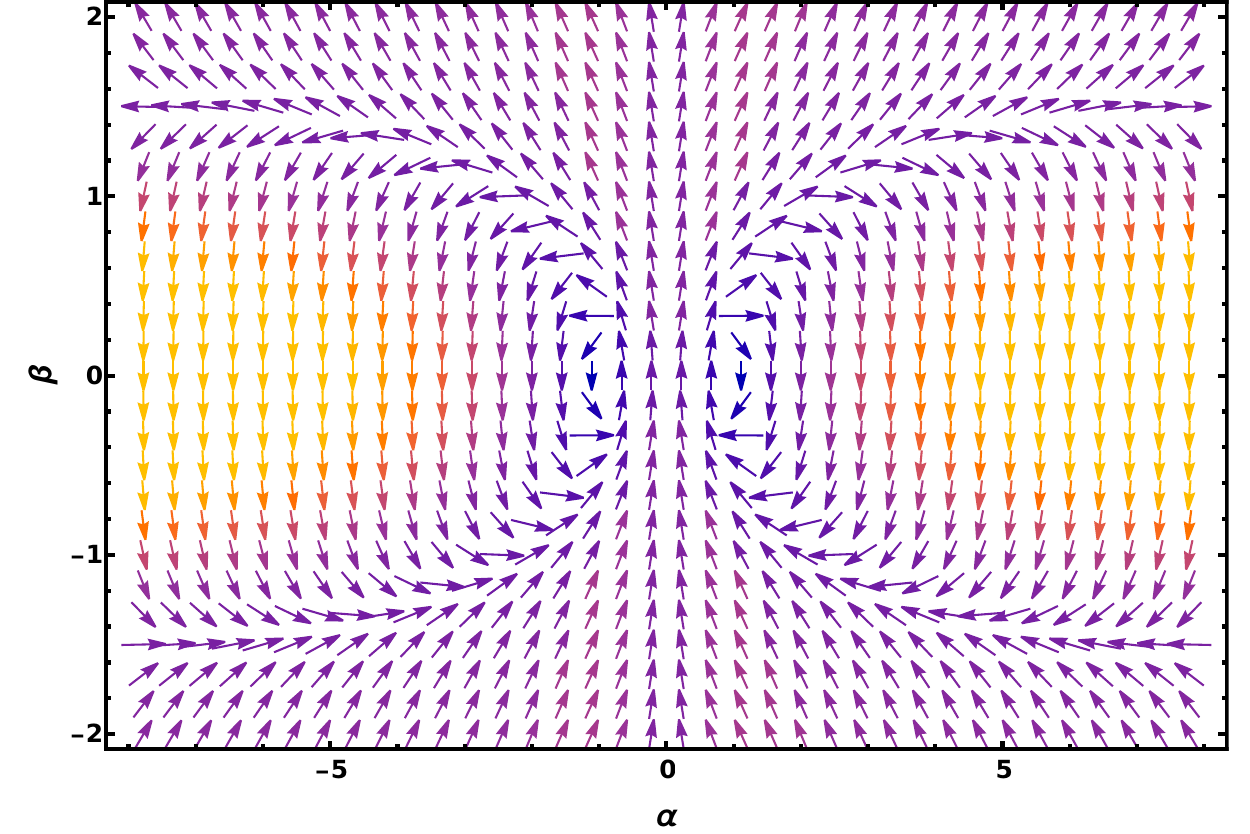}}\hfill%
     \end{center}
     \caption{Field plot of the guidance equations~\ref{dadtL},~\ref{dbdtL} showing Bohmian trajectories for the Lorentzian superposition with $\gamma=1$ and $k_0=0.1$. In both regions of the configuration space, closed-loop trajectories extend over substantial volume ranges, indicating quantum bounce behavior. The trajectories exhibit sharp turns due to the presence of the modulus $|\alpha \pm \beta|$ in the wavefunction $\Psi$. The color gradient highlights regions of rapid velocity flow and the rapid $\beta \rightarrow -\beta$ transition in anisotropy at approximately constant volumes..}
    \label{lorentzian-vplot}
\end{figure}

Fig.~\ref{lorentzian-vplot} shows a detailed vector field plot of the guidance equations~\ref{dadtL},~\ref{dbdtL} showing Bohmian trajectories where we specifically fix the parameters as $\gamma = 1$ and $k_0 = 0.1$ to reflect a wavepacket that is sufficiently localized with a slight shift in momentum space. The key feature is the presence of organized closed-loop trajectories forming symmetric ring-like structures around the line $\alpha=0$ in the $\alpha$-$\beta$ plane. A representative left-sector cycle ($\alpha<0$) begins with small volume (large negative $\alpha$, $\beta<0$), and expands with anisotropy reduction followed by contraction while anisotropy grows oppositely ($\beta>0$). At a constant minimum volume, rapid $\beta\to-\beta$ reversal occurs (high-velocity region shown by colour gradient), completing the cyclic evolution. These closed loop structures signify quantum bounces wherein the universe avoids the classical singularity by confining within a finite volume region, a signature of singularity resolution within Bohmian dynamics guided by the Lorentzian wavepacket.
\FloatBarrier
In comparison to Gaussian superpositions previously studied, the Lorentzian wavepacket produces a more structured and richer velocity field that exhibits these closed orbits and cusp-induced sharp deviations. The Gaussian case typically yields laminar diagonal flows with nearly parallel velocities and less effective singularity avoidance, due to the absence of strong interference structures between the high-$k$ modes of the left- and right-moving components of the wavefunction, unlike in the Lorentzian superposition~\ref{lorentzian-wavefunction}. The contrast emphasizes that the form of wavepacket superposition strongly governs the Bohmian dynamics. Specifically, the ability of the quantum gravitational framework to resolve the singularity fundamentally depends on the choice of $F(k)$ shaping the wavefunction.

\section{Non-Equilibrium Quantum Mechanics and Relaxation in Bohmian Cosmology}
\label{sec4}
In standard quantum mechanics, the Born rule---identifying the probability density $\rho(\mathbf{q},t)$ with $|\psi(\mathbf{q},t)|^2$---is a fundamental postulate. By contrast, de Broglie--Bohm pilot-wave theory treats this identification as an emergent equilibrium hypothesis rather than an axiom.

David Bohm already recognized the possibility of quantum non-equilibrium states $\rho\neq|\psi|^2$ in his seminal papers, noting that they could arise from modified guidance relations or additional forces in the particle equations of motion~\cite{bohm1952suggested,bohm1953proof}. Bohm and Vigier later explored related ideas, introducing irregular fluctuations in a fluid model to demonstrate relaxation of arbitrary distributions toward $|\psi|^2$, though these approaches remained conceptually incomplete~\cite{bohm1954model}.

Vigier--Bohm stochastic models lacked the chaos-based fine-graining required for quantum relaxation. In 1990, Antony Valentini proposed the subquantum $H$-theorem, establishing a dynamical basis for relaxation toward quantum equilibrium in Bohmian mechanics~\cite{valentini1991signal1,valentini1991signal2}. According to this theorem, non-equilibrium distributions evolve via the continuity equation $ \partial_t \rho + \nabla \cdot (\rho \mathbf{v}) = 0 $, with unmodified Bohmian velocities $\mathbf{v}=\nabla S/m$. For complex wavefunctions exhibiting chaotic streamline structure in configuration space, such distributions relax toward quantum equilibrium $\rho\to|\psi|^2$ through a process analogous to classical statistical mechanics relaxation. Valentini quantified this via the coarse-grained $H$-function,
\begin{equation}
\bar{H}(t) = \int d\mathbf{q} \, \bar{\rho}(\mathbf{q},t) \ln \left[ \frac{\bar{\rho}(\mathbf{q},t)}{|\psi(\mathbf{q},t)|^2} \right],
\end{equation}
where overlines denote averaging over finite cells. Its exponential decay, $\bar{H}(t) \to 0$, signals approach to the Born rule driven by irreversible fine-graining along Bohmian trajectories.

Valentini's $H$-theorem, supported by extensive numerical simulations, establishes quantum relaxation as a generic feature of complex quantum systems with chaotic configuration-space flows, analogous to the Boltzmann $H$-theorem for classical gases. Early works demonstrated rapid equilibration in simple systems such as the two-dimensional infinite square well potential, where relaxation timescales scale inversely with wavefunction complexity~\cite{valentini2005dynamical,towler2012time}. Subsequent studies of coupled harmonic oscillators revealed incomplete relaxation, depends on the coupling strength and interaction parameters~\cite{lustosa2021quantum,lustosa2023evolution}.

In cosmology, with quantum mechanical inflaton field living in the Friedman background, similar non-equilibrium formulation gains particular significance due to the extreme conditions of the early universe. Colin and Valentini demonstrated that primordial non-equilibrium---arising from incomplete relaxation during pre-inflationary phases---serves as a source of cosmic microwave background (CMB) anomalies, including large-scale power suppression and hemispherical asymmetry, producing observable deviations from standard inflationary predictions~\cite{valentini2010inflationary,colin2015primordial,colin2016robust}. Building on this foundation, Valentini argued that quantum gravity at the Planck scale---characterized by non-normalizable Wheeler-DeWitt wavefunctional---fundamentally lacks the Born rule, with relaxation emerging only during subsequent semiclassical evolution~\cite{valentini2023beyond}. This motivates our investigation of quantum relaxation within the Wheeler-DeWitt framework of quantum cosmology.

Our analysis of Gaussian and Lorentzian superpositions systematically tests this hypothesis, revealing how initial quantum non-equilibrium evolves under distinct Bohmian flows and whether relaxation dynamics correlate with cosmological singularity resolution.

\subsection{Relaxation Dynamics: Gaussian Superposition}
In this section, we systematically investigate non-equilibrium relaxation for Gaussian superposition wavepackets by numerically computing the $H$-function evolution. We have already computed Bohmian trajectories for the {\it equilibrium} case from $\Psi_G(\sigma=1,k_0=1)$, shown in Fig.~\ref{fig:gaussian-vplot}, which satisfies the Born rule $\rho_0=|\Psi_G|^2$ by construction. We also initialize a non-equilibrium distribution $\rho$ using a distinct Gaussian profile $\Psi_G(\sigma=0.5,k_0=2)$ and evolve it under the guidance equations~\ref{dalphadt},~\ref{dbetadt}.

\begin{figure}[ht!]
    \begin{center}
    
    \subfigure[]{\includegraphics[width=0.45 \textwidth,height= 5 cm]{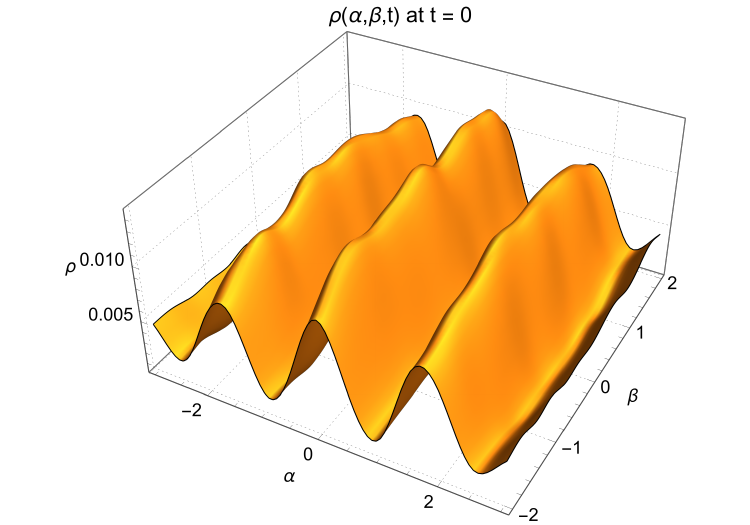}}\hfill%
    \subfigure[]{\includegraphics[width=0.45 \textwidth,height= 5 cm]{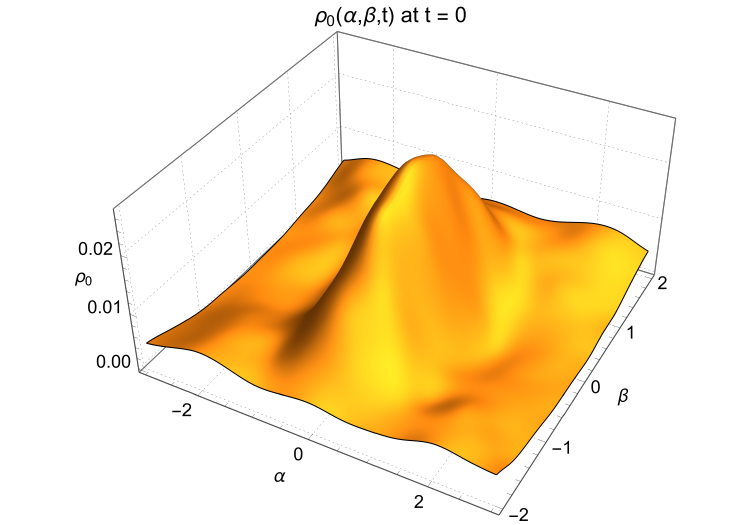}}\hfill%

    \subfigure[]{\includegraphics[width=0.45 \textwidth,height= 5 cm]{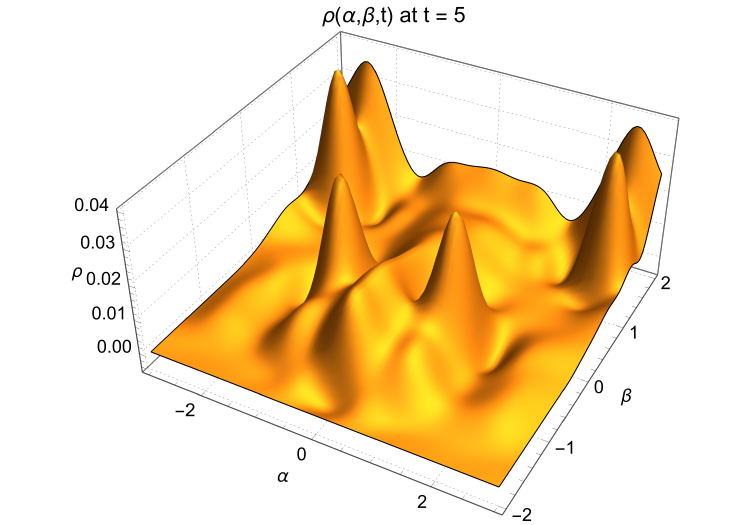}}\hfill%
    \subfigure[]{\includegraphics[width=0.45 \textwidth,height= 5 cm]{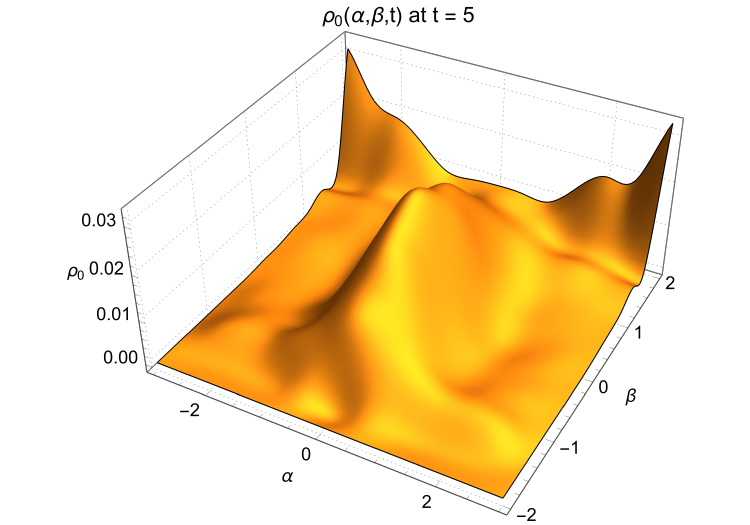}}\hfill%
   \subfigure[]{\includegraphics[width=0.45 \textwidth,height= 5 cm]{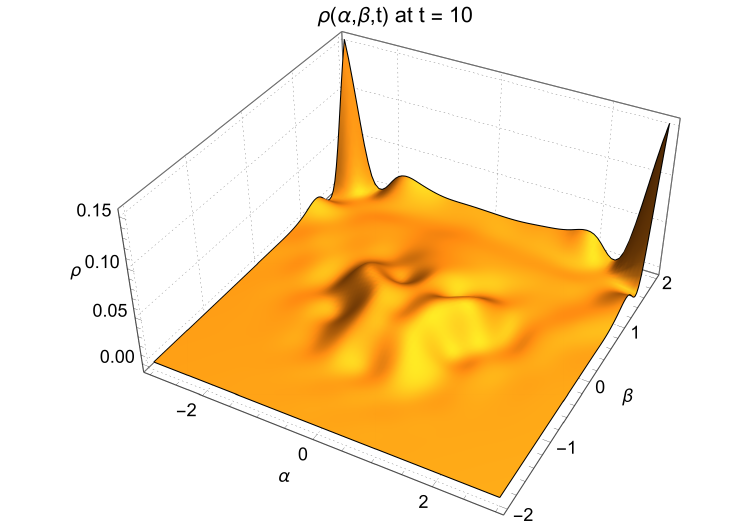}}\hfill%
    \subfigure[]{\includegraphics[width=0.45 \textwidth,height= 5 cm]{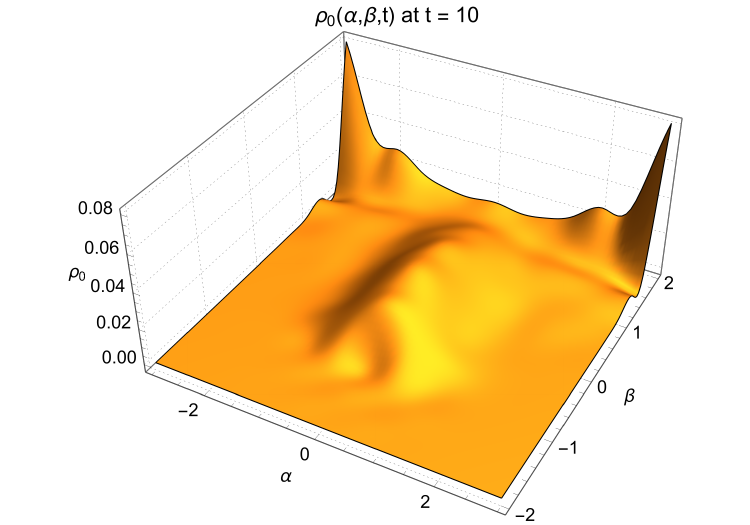}}\hfill%
    \end{center}
    \caption{The left and right columns show the evolution of the non-equilibrium distribution $\rho(\alpha,\beta)=|\Psi_G(\sigma=0.5,k_0=2)|^2$ and the equilibrium distribution $\rho_0(\alpha,\beta)=|\Psi_G(\sigma=1,k_0=1)|^2$, respectively, under the Bohmian flow of the Gaussian superposition case~\ref{fig:gaussian-vplot} over the time interval $t \in [0,10]$. Both distributions appear suppressed at $t=10$ due to the accumulation of sample points near the boundaries caused by the laminar flow. While the equilibrium distribution largely preserves its overall shape under the Bohmian flow, the non-equilibrium distribution leaves residual structures near the closed-orbit regions.}
    \label{fig:Gaussian_evolution}
\end{figure}

\begin{figure}[ht!]
    \begin{center}
    {\includegraphics[width=0.75 \textwidth,height =7 cm]{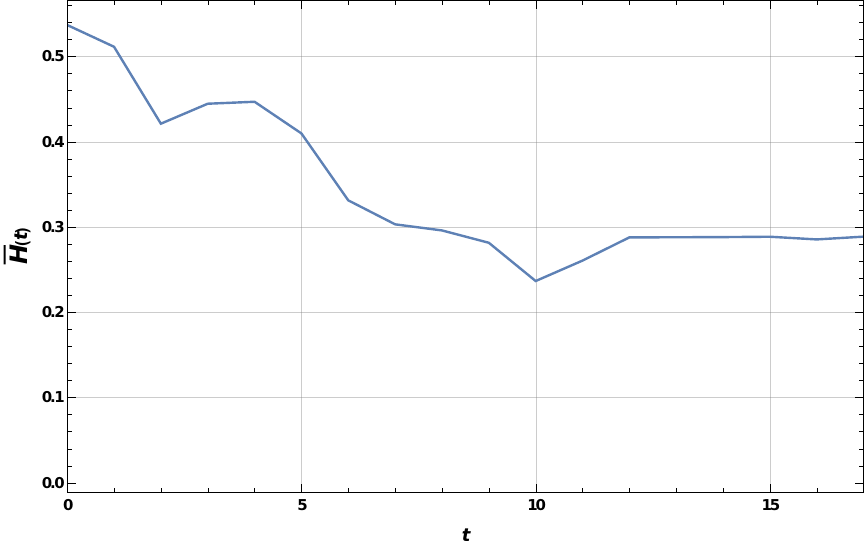}}\hfill%
     \end{center}
     \caption{Time evolution of the $H$-function for the Gaussian superposition, computed within the central region while excluding sample points that reach the boundaries during the evolution. The $H$-function exhibits a non-monotonic decrease followed by a plateau starting at $t \approx 10$, indicating saturation of the relaxation process.}
    \label{fig:Hplot_G}
\end{figure}

Fig.~\ref{fig:Gaussian_evolution} illustrates these dynamics. The Bohmian velocity field for the Gaussian superposition, shown in Fig.~\ref{fig:gaussian-vplot}, exhibits laminar diagonal streamlines converging toward configuration space boundaries. Under equilibrium initial conditions $\rho_0(\alpha,\beta)=|\Psi_G(\sigma=1,k_0=1)|^2$, sample points trace these streamlines with characteristic accumulation at diagonal corners while preserving the overall equilibrium distribution shape. For the non-equilibrium distribution $\rho(\alpha,\beta)=|\Psi_G(\sigma=0.5,k_0=2)|^2$, streamline advection similarly drives universes toward boundaries, leaving residual populations near $\alpha\approx0$ (volume transition region) and localized closed-orbit zones. The final non-equilibrium distribution remains distinctly non-Bornian, exhibiting pronounced boundary accumulation rather than relaxation toward equilibrium. This incomplete relaxation becomes clearly evident when computing Valentini's coarse-grained $H$-function,
\begin{equation}
\bar{H}(t) = \int d\alpha\, d\beta \, \bar{\rho}(\alpha,\beta,t) \ln \left[ \frac{\bar{\rho}(\alpha,\beta,t)}{|\Psi_G(\alpha,\beta)|^2} \right].
\label{coarseH}
\end{equation}
\FloatBarrier
The time evolution of the $H$-function is shown in Fig.~\ref{fig:Hplot_G}. The difference in the two distributions determining the numerator and denominator (logarthmic argument in eq. \ref{coarseH}) gives the initial value of $H$-function different from zero, $\bar{H}(0) \approx 0.5$. Contrary to the monotonic decay characteristic of chaotic Bohmian flows, $\bar{H}(t)$ exhibits only weak, non-monotonic decay over $t \in [0,10]$, followed by a plateau after $t \approx 10$, without approaching the equilibrium value $\bar{H}(t) \to 0$. This saturation of the $H$-function indicates that mixing ceases and quantum relaxation halts short of equilibrium.  
\FloatBarrier
Having established that Gaussian superpositions produce non-chaotic  Bohmian flows insufficient for complete relaxation in minisuperspace, we now test the Lorentzian case. Unlike the Gaussian velocity field, the Lorentzian superposition generates a closed-loop flow structure (Fig.~\ref{lorentzian-vplot}) with enhanced streamline complexity featuring organized bounce orbits. This structure may promote superior mixing through repeated particle circulation and scattering, potentially providing the chaotic fine-graining necessary for relaxation.

\subsection{Relaxation Dynamics: Lorentzian Superposition}
Following the Gaussian analysis, we now examine quantum relaxation for Lorentzian superposition wave packets. The equilibrium guiding field is provided by $\Psi(\gamma=1,k_0=0.1)$, satisfying $\rho_0=|\Psi|^2$ and generating the Bohmian flow shown in Fig~\ref{lorentzian-vplot}. We initialize a non-equilibrium distribution $\rho(\alpha,\beta) = |\Psi(\gamma=0.05,k_0 = 0.5)|^2$ and evolve it under this guidance field, testing whether the Lorentzian flow's characteristic closed-loop trajectories drives convergence toward equilibrium through dynamical mixing.

\begin{figure}[ht!]
    \begin{center}
    
    \subfigure[]{\includegraphics[width=0.45 \textwidth,height= 5 cm]{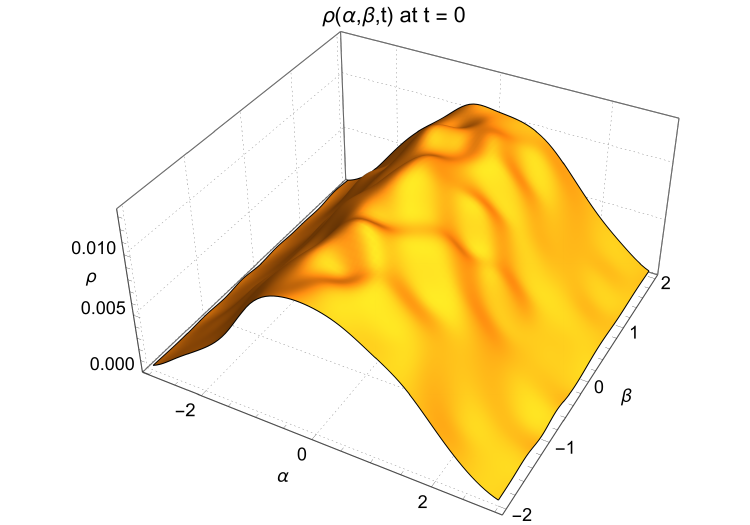}}\hfill%
    \subfigure[]{\includegraphics[width=0.45 \textwidth,height= 5 cm]{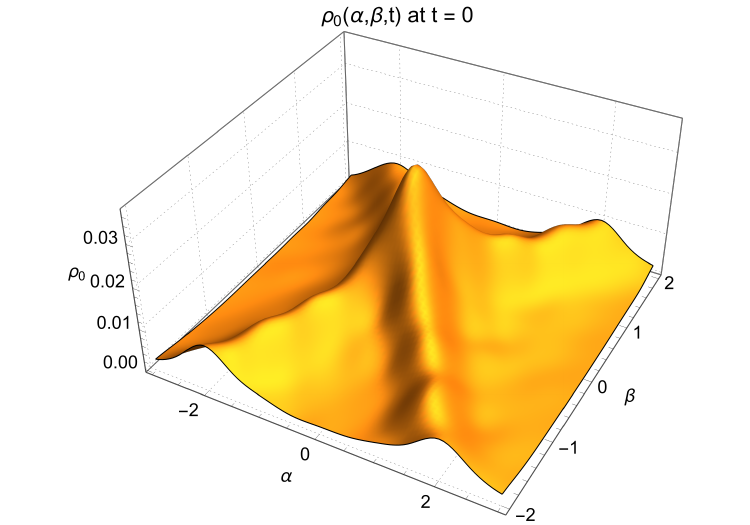}}\hfill%

    \subfigure[]{\includegraphics[width=0.45 \textwidth,height= 5 cm]{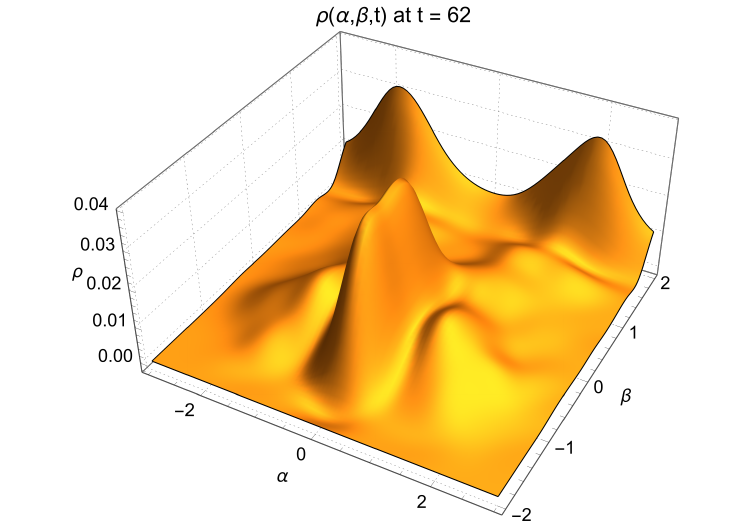}}\hfill%
    \subfigure[]{\includegraphics[width=0.45 \textwidth,height= 5 cm]{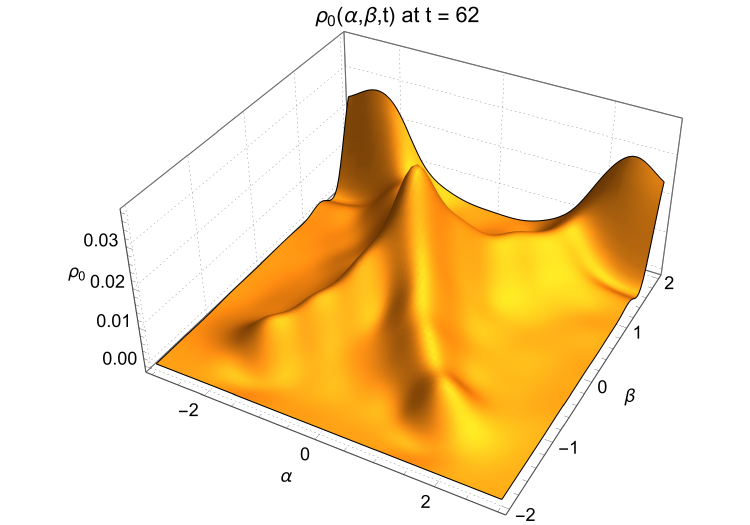}}\hfill%
   \subfigure[]{\includegraphics[width=0.45 \textwidth,height= 5 cm]{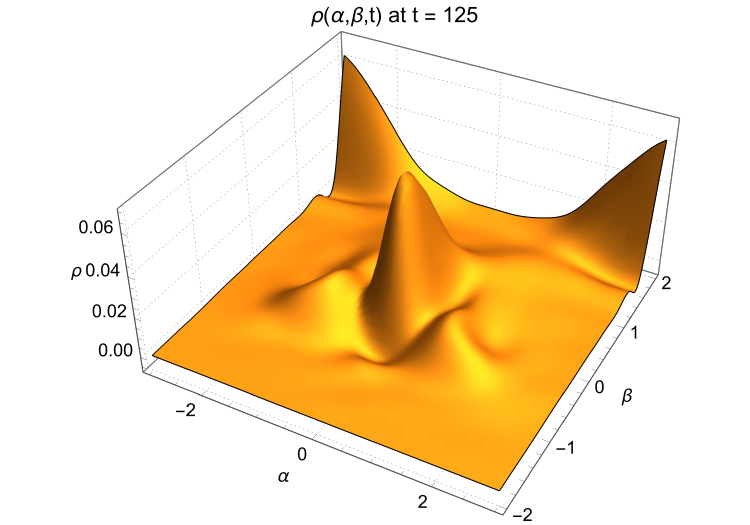}}\hfill%
    \subfigure[]{\includegraphics[width=0.45 \textwidth,height= 5 cm]{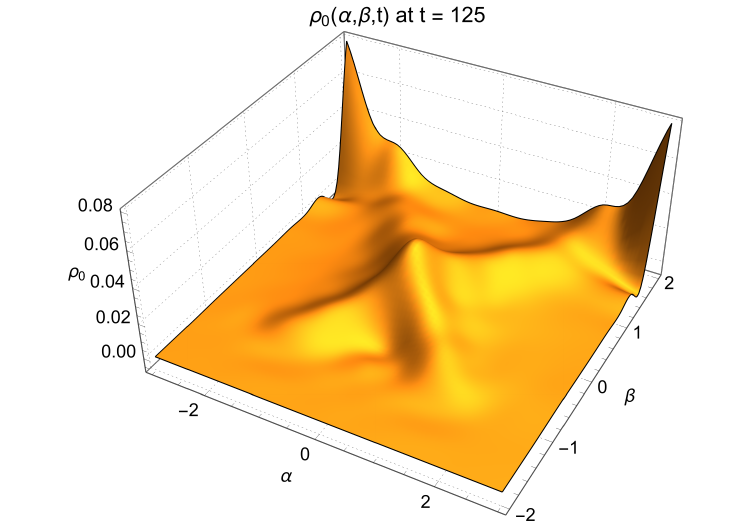}}\hfill%
    \end{center}
    \caption{The left and right columns show the evolution of the non-equilibrium distribution $\rho(\alpha,\beta)=|\Psi(\gamma=1,k_0=0.1)|^2$ and the equilibrium distribution $\rho_0(\alpha,\beta)=|\Psi(\gamma=0.05,k_0=0.5)|^2$, respectively, under the Bohmian flow of the Lorentzian superposition case~\ref{lorentzian-vplot} over the time interval $t \in [0,125]$. The equilibrium distribution remains stationary, while the non-equilibrium distribution develops partial structural similarity to the equilibrium distribution by $t=125$. During the evolution, sample points accumulate near the boundaries in both distributions.
}
    \label{fig:Lorentzian_evolution}
\end{figure}

\begin{figure}[ht!]
    \begin{center}
    {\includegraphics[width=0.75 \textwidth,height =7 cm]{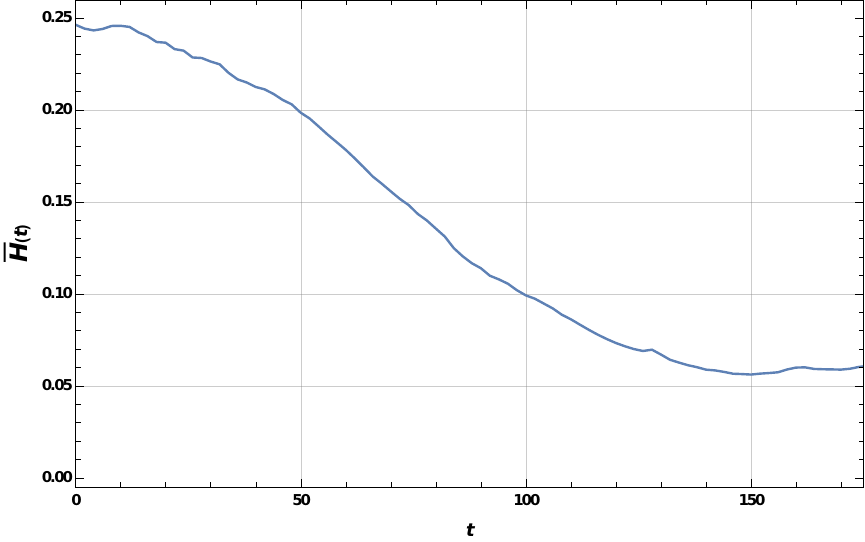}}\hfill%
     \end{center}
     \caption{Time evolution of the $H$-function for the Lorentzian superposition, computed within the central region while excluding sample points that reach the boundaries during the evolution. The $H$-function exhibits a monotonic decrease followed by a plateau starting at $t \approx 125$, indicating saturation of the relaxation process.}
    \label{fig:Hplot-L}
\end{figure}

Fig.~\ref{fig:Lorentzian_evolution} compares the evolution of non-equilibrium $\rho(\alpha,\beta)=|\Psi(\gamma=1,k_0=0.1)|^2$ and equilibrium $\rho_0(\alpha,\beta)=|\Psi(\gamma=0.05,k_0=0.5)|^2$ distributions under the Lorentzian velocity field, which features organized bounce orbits characteristic of singularity avoidance. The equilibrium distribution remains stationary under its self-consistent guidance field, exhibiting characteristic boundary accumulation from outward streamline flow. In contrast, the non-equilibrium distribution undergoes significant evolution, developing partial structural similarity to the equilibrium ---indicative of mixing-driven relaxation toward Born quantum equilibrium.
\FloatBarrier
The incomplete relaxation in the Lorentzian case is quantified in Fig.~\ref{fig:Hplot-L}. The coarse-grained $H$-function decays from $\bar{H}(0)\approx0.25$ to $\bar{H}(125)\approx0.05$, exhibiting monotonic evolution until $t\approx 125$ and then saturation. While this demonstrates substantially improved mixing relative to the Gaussian case due to enhanced streamline complexity, the saturation at $\bar{H}(125)\approx0.05$ reveals that the structured closed-loop flow remains insufficiently chaotic to drive complete equilibration.
\FloatBarrier

\section{Discussion and Conclusion}
\label{sec5}
In this paper, we investigated the Big Bang singularity resolution and quantum relaxation within Bohmian quantum cosmology, employing a plane-symmetric Bianchi type-I minisuperspace model in the Wheeler-DeWitt framework. Starting from the minisuperspace model, we derived the Wheeler-DeWitt equation and obtained physically motivated wavefunctions as superpositions of left- and right-moving plane-wave modes in the $(\alpha,\beta)$ volume-anisotropy configuration space. Gaussian and Lorentzian wavepackets provided systematic test cases for analyzing both trajectory dynamics and quantum relaxation toward the Born rule.

In the Gaussian superposition case, the majority of Bohmian trajectories exhibit classical behavior, manifesting both past and future singularities. Only a negligible fraction forms small-amplitude oscillating (cyclic) universes confined to regions of significant quantum potential repulsion, with spatial extents far smaller than observationally relevant cosmological scales. Unlike Gaussian wave packets, where high-$k$ components decay exponentially, the Lorentzian superposition characterizes long tails sustaining higher-momentum $k$-modes that extend into classically forbidden regions. This generates stronger quantum potential barriers near the singularity, producing a more structured velocity field with closed, non-singular bounce trajectories spanning substantial volume ranges. While some singular trajectories persist, their fraction is substantially reduced compared to the Gaussian case. This demonstrates superior singularity resolution with Lorentzian superposition, reflecting the enhanced role of power-law momentum tails and resulting in nodal complexity due to the modulus $|\alpha\pm\beta|$  occuring in the wavefunction.

Building on the trajectory analysis, we next examined quantum relaxation under the guidance fields of these two types of wavefunctions. For the Gaussian superposition $\Psi_G(\sigma=1,d=1)$, laminar diagonal streamlines rapidly advect the initial non-equilibrium  distribution $\rho_0=|\Psi_G(\sigma=0.5,d=2)|^2$ towards configuration space boundaries. This boundary pile-up leaves only residual sample populations near localized closed-orbit zones, yielding non-monotonic decay of the $H$-function from $0.5\to0.3$ over $t\in[0,10]$, followed by saturation at $\bar{H}(t)\approx0.3$ for $t>10$. In contrast,
the Lorentzian superposition wavepacket $\Psi(\gamma=1.0,k_0 = 0.1)$ exhibits substantially richer relaxation dynamics. While the non-equilibrium distribution $\rho = |\Psi(\gamma=0.05,k_0 = 0.5)|^2$, develops partial structural similarity to the equilibrium $|\Psi|^2$—particularly in central regions—it retains the pronounced boundary pile-up similar to the laminar Gaussian flow, preventing complete Born rule relaxation. This dynamics manifests as monotonic decay of the $H$-function from $0.25 \to 0.05$ over the extended interval $t\in[0,125]$ followed by saturation at $\bar{H}(t)\approx0.05$ for $t>125$.

Thus, the Lorentzian profile outperforms the Gaussian envelope across both situations: complex closed-loop bounce trajectories simultaneously enable non-singular evolution and drive monotonic $H$-function decay via chaotic mixing, whereas Gaussian superposition yield markedly poor relaxation with substantial singular trajectories.

Our analysis thus reveals that the type of wave packet superposition fundamentally governs both singularity resolution and quantum relaxation through Bohmian flow structure. The resulting velocity field geometry—ranging from laminar streamlines to chaotic mixing—directly determines both the fraction of non-singular bounce trajectories and the rate of $H$-function decay toward Born-rule equilibration. These dual criteria reveal that optimal singularity resolution requires the same flow complexity that drives effective quantum relaxation.

Such control by wavefunction structure has profound implications for quantum gravity. The persistent incomplete relaxation observed across both cases—even with Lorentzian superposition—establishes primordial statistical relics via Born rule deviations surviving from Planck scales. Thus, quantum nonequilibrium persists during primordial evolution, with full equilibration emerging only during subsequent semiclassical expansion toward classical cosmology.

These findings position pilot-wave theory as a predictive framework for quantum cosmology, where observable relics of primordial non-equilibrium would provide empirical tests. Future work should explore more complex multi-mode superpositions that may achieve the sufficient streamline chaos required for complete $\bar{H}(t)\to0$ relaxation while maintaining robust singularity resolution. Extending beyond minisuperspace by incorporating matter fields or higher-dimensional configuration spaces could generate the enhanced trajectory complexity necessary to test full equilibration in more complex cosmological models.

\section{Acknowledgements}
Vishal is supported by a research fellowship from the Ministry of Human Resource Development (MHRD), Government of India. The authors would like to thank the Indian Institute of Technology Guwahati for providing computing and supercomputing facilities.


\begin{thebibliography}{10}

\bibitem{friedman1922krummung}
Alexander Friedman.
\newblock {\"U}ber die kr{\"u}mmung des raumes.
\newblock {\em Zeitschrift f{\"u}r Physik}, 10(1):377--386, 1922.

\bibitem{lemaitre1927univers}
G.~{Lema{\^\i}tre}.
\newblock {Un Univers homog{\`e}ne de masse constante et de rayon croissant
  rendant compte de la vitesse radiale des n{\'e}buleuses extra-galactiques}.
\newblock {\em Annales de la Soci{\'e}t{\'e} Scientifique de Bruxelles},
  47:49--59, January 1927.

\bibitem{robertson1935kinematics}
H.~P. {Robertson}.
\newblock {Kinematics and World-Structure}.
\newblock {\em \apj}, 82:284, November 1935.
\newblock \href {https://doi.org/10.1086/143681} {\path{doi:10.1086/143681}}.

\bibitem{walker1937milne}
A.~G. Walker.
\newblock On milne's theory of world-structure.
\newblock {\em Proceedings of the London Mathematical Society},
  s2-42(1):90--127, 1937.
\newblock URL:
  \url{https://londmathsoc.onlinelibrary.wiley.com/doi/abs/10.1112/plms/s2-42.1.90},
  \href
  {https://arxiv.org/abs/https://londmathsoc.onlinelibrary.wiley.com/doi/pdf/10.1112/plms/s2-42.1.90}
  {\path{arXiv:https://londmathsoc.onlinelibrary.wiley.com/doi/pdf/10.1112/plms/s2-42.1.90}},
  \href {https://doi.org/10.1112/plms/s2-42.1.90}
  {\path{doi:10.1112/plms/s2-42.1.90}}.

\bibitem{raychaudhuri1955relativistic}
Amalkumar Raychaudhuri.
\newblock Relativistic cosmology. i.
\newblock {\em Phys. Rev.}, 98:1123--1126, May 1955.
\newblock URL: \url{https://link.aps.org/doi/10.1103/PhysRev.98.1123}, \href
  {https://doi.org/10.1103/PhysRev.98.1123}
  {\path{doi:10.1103/PhysRev.98.1123}}.

\bibitem{landau1971classical}
LD~Landau and EM~Lifshitz.
\newblock {\em The classical theory of fields, course of theoretical physics}.
\newblock 1971.

\bibitem{ryan2015homogeneous}
Michael~P. Ryan and Lawrence~C. Shepley.
\newblock {\em {Homogeneous Relativistic Cosmologies}}.
\newblock Princeton Series in Physics. Princeton University Press, Princeton,
  1975.

\bibitem{ellis2006bianchi}
G.~F.~R. {Ellis}.
\newblock {The Bianchi models: Then and now}.
\newblock {\em General Relativity and Gravitation}, 38(6):1003--1015, June
  2006.
\newblock \href {https://doi.org/10.1007/s10714-006-0283-4}
  {\path{doi:10.1007/s10714-006-0283-4}}.

\bibitem{collins1979singularities}
C.B. Collins and G.F.R. Ellis.
\newblock Singularities in bianchi cosmologies.
\newblock {\em Physics Reports}, 56(2):65--105, 1979.
\newblock URL:
  \url{https://www.sciencedirect.com/science/article/pii/0370157379900656},
  \href {https://doi.org/10.1016/0370-1573(79)90065-6}
  {\path{doi:10.1016/0370-1573(79)90065-6}}.

\bibitem{ellis1999cosmological}
George F.~R. Ellis and Henk van Elst.
\newblock {Cosmological models: Cargese lectures 1998}.
\newblock {\em NATO Sci. Ser. C}, 541:1--116, 1999.
\newblock \href {https://arxiv.org/abs/gr-qc/9812046}
  {\path{arXiv:gr-qc/9812046}}, \href
  {https://doi.org/10.1007/978-94-011-4455-1_1}
  {\path{doi:10.1007/978-94-011-4455-1_1}}.

\bibitem{wheeler1969superspace}
J.~A. Wheeler.
\newblock {SUPERSPACE AND THE NATURE OF QUANTUM GEOMETRODYNAMICS}.
\newblock {\em Adv. Ser. Astrophys. Cosmol.}, 3:27--92, 1987.

\bibitem{dewitt1967quantum}
Bryce~S. DeWitt.
\newblock Quantum theory of gravity. i. the canonical theory.
\newblock {\em Phys. Rev.}, 160:1113--1148, Aug 1967.
\newblock URL: \url{https://link.aps.org/doi/10.1103/PhysRev.160.1113}, \href
  {https://doi.org/10.1103/PhysRev.160.1113}
  {\path{doi:10.1103/PhysRev.160.1113}}.

\bibitem{hartle1983wave}
J.~B. Hartle and S.~W. Hawking.
\newblock Wave function of the universe.
\newblock {\em Phys. Rev. D}, 28:2960--2975, Dec 1983.
\newblock URL: \url{https://link.aps.org/doi/10.1103/PhysRevD.28.2960}, \href
  {https://doi.org/10.1103/PhysRevD.28.2960}
  {\path{doi:10.1103/PhysRevD.28.2960}}.

\bibitem{misner1969quantum}
Charles~W. Misner.
\newblock Quantum cosmology. i.
\newblock {\em Phys. Rev.}, 186:1319--1327, Oct 1969.
\newblock URL: \url{https://link.aps.org/doi/10.1103/PhysRev.186.1319}, \href
  {https://doi.org/10.1103/PhysRev.186.1319}
  {\path{doi:10.1103/PhysRev.186.1319}}.

\bibitem{kiefer2019singularity}
Claus Kiefer, Nick Kwidzinski, and Dennis Piontek.
\newblock {Singularity avoidance in Bianchi I quantum cosmology}.
\newblock {\em Eur. Phys. J. C}, 79(8):686, 2019.
\newblock \href {https://arxiv.org/abs/1903.04391} {\path{arXiv:1903.04391}},
  \href {https://doi.org/10.1140/epjc/s10052-019-7193-6}
  {\path{doi:10.1140/epjc/s10052-019-7193-6}}.

\bibitem{bouhmadi2014resolution}
Mariam Bouhmadi-L\'opez, Claus Kiefer, and Manuel Kr\"amer.
\newblock Resolution of type iv singularities in quantum cosmology.
\newblock {\em Phys. Rev. D}, 89:064016, Mar 2014.
\newblock URL: \url{https://link.aps.org/doi/10.1103/PhysRevD.89.064016}, \href
  {https://doi.org/10.1103/PhysRevD.89.064016}
  {\path{doi:10.1103/PhysRevD.89.064016}}.

\bibitem{kuchavr2011time}
K.~V. Kuchar.
\newblock {Time and interpretations of quantum gravity}.
\newblock {\em Int. J. Mod. Phys. D}, 20:3--86, 2011.
\newblock \href {https://doi.org/10.1142/S0218271811019347}
  {\path{doi:10.1142/S0218271811019347}}.

\bibitem{isham1993canonical}
C.~J. Isham.
\newblock {Canonical quantum gravity and the problem of time}.
\newblock {\em NATO Sci. Ser. C}, 409:157--287, 1993.
\newblock \href {https://arxiv.org/abs/gr-qc/9210011}
  {\path{arXiv:gr-qc/9210011}}.

\bibitem{wald1980quantum}
Robert~M. Wald.
\newblock Quantum gravity and time reversibility.
\newblock {\em Phys. Rev. D}, 21:2742--2755, May 1980.
\newblock URL: \url{https://link.aps.org/doi/10.1103/PhysRevD.21.2742}, \href
  {https://doi.org/10.1103/PhysRevD.21.2742}
  {\path{doi:10.1103/PhysRevD.21.2742}}.

\bibitem{unruh1989time}
William~G. Unruh and Robert~M. Wald.
\newblock Time and the interpretation of canonical quantum gravity.
\newblock {\em Phys. Rev. D}, 40:2598--2614, Oct 1989.
\newblock URL: \url{https://link.aps.org/doi/10.1103/PhysRevD.40.2598}, \href
  {https://doi.org/10.1103/PhysRevD.40.2598}
  {\path{doi:10.1103/PhysRevD.40.2598}}.

\bibitem{gambini2009conditional}
Rodolfo Gambini, Rafael~A. Porto, Jorge Pullin, and Sebasti\'an Torterolo.
\newblock Conditional probabilities with dirac observables and the problem of
  time in quantum gravity.
\newblock {\em Phys. Rev. D}, 79:041501, Feb 2009.
\newblock URL: \url{https://link.aps.org/doi/10.1103/PhysRevD.79.041501}, \href
  {https://doi.org/10.1103/PhysRevD.79.041501}
  {\path{doi:10.1103/PhysRevD.79.041501}}.

\bibitem{anderson2012problem}
E.~Anderson.
\newblock Problem of time in quantum gravity.
\newblock {\em Annalen der Physik}, 524(12):757--786, 2012.
\newblock URL:
  \url{https://onlinelibrary.wiley.com/doi/abs/10.1002/andp.201200147}, \href
  {https://arxiv.org/abs/https://onlinelibrary.wiley.com/doi/pdf/10.1002/andp.201200147}
  {\path{arXiv:https://onlinelibrary.wiley.com/doi/pdf/10.1002/andp.201200147}},
  \href {https://doi.org/10.1002/andp.201200147}
  {\path{doi:10.1002/andp.201200147}}.

\bibitem{malkiewicz2020quantum}
Przemys\l{}aw Ma\l{}kiewicz, Patrick Peter, and S.~D.~P. Vitenti.
\newblock Quantum empty bianchi i spacetime with internal time.
\newblock {\em Phys. Rev. D}, 101:046012, Feb 2020.
\newblock URL: \url{https://link.aps.org/doi/10.1103/PhysRevD.101.046012},
  \href {https://doi.org/10.1103/PhysRevD.101.046012}
  {\path{doi:10.1103/PhysRevD.101.046012}}.

\bibitem{hohn2021trinity}
Philipp~A. H\"ohn, Alexander R.~H. Smith, and Maximilian P.~E. Lock.
\newblock Trinity of relational quantum dynamics.
\newblock {\em Phys. Rev. D}, 104:066001, Sep 2021.
\newblock URL: \url{https://link.aps.org/doi/10.1103/PhysRevD.104.066001},
  \href {https://doi.org/10.1103/PhysRevD.104.066001}
  {\path{doi:10.1103/PhysRevD.104.066001}}.

\bibitem{rovelli1991time}
Carlo Rovelli.
\newblock Time in quantum gravity: An hypothesis.
\newblock {\em Phys. Rev. D}, 43:442--456, Jan 1991.
\newblock URL: \url{https://link.aps.org/doi/10.1103/PhysRevD.43.442}, \href
  {https://doi.org/10.1103/PhysRevD.43.442}
  {\path{doi:10.1103/PhysRevD.43.442}}.

\bibitem{page1983evolution}
Don~N. Page and William~K. Wootters.
\newblock Evolution without evolution: Dynamics described by stationary
  observables.
\newblock {\em Phys. Rev. D}, 27:2885--2892, Jun 1983.
\newblock URL: \url{https://link.aps.org/doi/10.1103/PhysRevD.27.2885}, \href
  {https://doi.org/10.1103/PhysRevD.27.2885}
  {\path{doi:10.1103/PhysRevD.27.2885}}.

\bibitem{bohm1952suggested}
David Bohm.
\newblock A suggested interpretation of the quantum theory in terms of "hidden"
  variables. i.
\newblock {\em Phys. Rev.}, 85:166--179, Jan 1952.
\newblock URL: \url{https://link.aps.org/doi/10.1103/PhysRev.85.166}, \href
  {https://doi.org/10.1103/PhysRev.85.166} {\path{doi:10.1103/PhysRev.85.166}}.

\bibitem{bohm2006undivided}
David Bohm and Basil~J Hiley.
\newblock {\em The undivided universe: An ontological interpretation of quantum
  theory}.
\newblock Routledge, 2006.

\bibitem{holland1995quantum}
Peter~R. Holland.
\newblock {\em The Quantum Theory of Motion: An Account of the de Broglie-Bohm
  Causal Interpretation of Quantum Mechanics}.
\newblock Cambridge University Press, 1993.

\bibitem{bohm1953proof}
David Bohm.
\newblock Proof that probability density approaches ${|\ensuremath{\psi}|}^{2}$
  in causal interpretation of the quantum theory.
\newblock {\em Phys. Rev.}, 89:458--466, Jan 1953.
\newblock URL: \url{https://link.aps.org/doi/10.1103/PhysRev.89.458}, \href
  {https://doi.org/10.1103/PhysRev.89.458} {\path{doi:10.1103/PhysRev.89.458}}.

\bibitem{colistete2000gaussian}
R.~Colistete, J.~C. Fabris, and N.~Pinto-Neto.
\newblock Gaussian superpositions in scalar-tensor quantum cosmological models.
\newblock {\em Phys. Rev. D}, 62:083507, Sep 2000.
\newblock URL: \url{https://link.aps.org/doi/10.1103/PhysRevD.62.083507}, \href
  {https://doi.org/10.1103/PhysRevD.62.083507}
  {\path{doi:10.1103/PhysRevD.62.083507}}.

\bibitem{alvarenga2002quantum}
F.~G. Alvarenga, J.~C. Fabris, N.~A. Lemos, and G.~A. Monerat.
\newblock Quantum cosmological perfect fluid models.
\newblock {\em General Relativity and Gravitation}, 34(5):651–663, May 2002.
\newblock URL: \url{http://dx.doi.org/10.1023/A:1015986011295}, \href
  {https://doi.org/10.1023/a:1015986011295}
  {\path{doi:10.1023/a:1015986011295}}.

\bibitem{pinto2013quantum}
N~Pinto-Neto and J~C Fabris.
\newblock Quantum cosmology from the de broglie–bohm perspective.
\newblock {\em Classical and Quantum Gravity}, 30(14):143001, jun 2013.
\newblock \href {https://doi.org/10.1088/0264-9381/30/14/143001}
  {\path{doi:10.1088/0264-9381/30/14/143001}}.

\bibitem{vicente2023bouncing}
G.~S. Vicente, Rudnei~O. Ramos, and Vit\'oria~N. Magalh\~aes.
\newblock Bouncing and inflationary dynamics in quantum cosmology in the de
  broglie--bohm interpretation.
\newblock {\em Phys. Rev. D}, 108:023517, Jul 2023.
\newblock URL: \url{https://link.aps.org/doi/10.1103/PhysRevD.108.023517},
  \href {https://doi.org/10.1103/PhysRevD.108.023517}
  {\path{doi:10.1103/PhysRevD.108.023517}}.

\bibitem{pinto2000quantum}
N.~Pinto-Neto, A.F. Velasco, and R.~Colistete.
\newblock Quantum isotropization of the universe.
\newblock {\em Physics Letters A}, 277(4):194--204, 2000.
\newblock URL:
  \url{https://www.sciencedirect.com/science/article/pii/S0375960100007064},
  \href {https://doi.org/10.1016/S0375-9601(00)00706-4}
  {\path{doi:10.1016/S0375-9601(00)00706-4}}.

\bibitem{tavakoli2019bianchi}
Fatimah Tavakoli and Babak Vakili.
\newblock Bianchi type i, schutz perfect fluid and evolutionary quantum
  cosmology.
\newblock {\em General Relativity and Gravitation}, 51(9), September 2019.
\newblock URL: \url{http://dx.doi.org/10.1007/s10714-019-2602-6}, \href
  {https://doi.org/10.1007/s10714-019-2602-6}
  {\path{doi:10.1007/s10714-019-2602-6}}.

\bibitem{valentini1991signal1}
Antony Valentini.
\newblock Signal-locality, uncertainty, and the subquantum h-theorem. i.
\newblock {\em Physics Letters A}, 156(1):5--11, 1991.
\newblock URL:
  \url{https://www.sciencedirect.com/science/article/pii/037596019190116P},
  \href {https://doi.org/10.1016/0375-9601(91)90116-P}
  {\path{doi:10.1016/0375-9601(91)90116-P}}.

\bibitem{valentini1991signal2}
Antony Valentini.
\newblock Signal-locality, uncertainty, and the subquantum h-theorem. ii.
\newblock {\em Physics Letters A}, 158(1):1--8, 1991.
\newblock URL:
  \url{https://www.sciencedirect.com/science/article/pii/037596019190330B},
  \href {https://doi.org/10.1016/0375-9601(91)90330-B}
  {\path{doi:10.1016/0375-9601(91)90330-B}}.

\bibitem{valentini2005dynamical}
Antony Valentini and Hans Westman.
\newblock Dynamical origin of quantum probabilities.
\newblock {\em Proceedings of the Royal Society A: Mathematical, Physical and
  Engineering Sciences}, 461(2053):253--272, 01 2005.
\newblock \href
  {https://arxiv.org/abs/https://royalsocietypublishing.org/rspa/article-pdf/461/2053/253/641588/rspa.2004.1394.pdf}
  {\path{arXiv:https://royalsocietypublishing.org/rspa/article-pdf/461/2053/253/641588/rspa.2004.1394.pdf}},
  \href {https://doi.org/10.1098/rspa.2004.1394}
  {\path{doi:10.1098/rspa.2004.1394}}.

\bibitem{towler2012time}
M.~D. Towler, N.~J. Russell, and Antony Valentini.
\newblock Time scales for dynamical relaxation to the born rule.
\newblock {\em Proceedings of the Royal Society A: Mathematical, Physical and
  Engineering Sciences}, 468(2140):990--1013, 11 2011.
\newblock \href
  {https://arxiv.org/abs/https://royalsocietypublishing.org/rspa/article-pdf/468/2140/990/813330/rspa.2011.0598.pdf}
  {\path{arXiv:https://royalsocietypublishing.org/rspa/article-pdf/468/2140/990/813330/rspa.2011.0598.pdf}},
  \href {https://doi.org/10.1098/rspa.2011.0598}
  {\path{doi:10.1098/rspa.2011.0598}}.

\bibitem{abraham2014long}
Eitan Abraham, Samuel Colin, and Antony Valentini.
\newblock Long-time relaxation in pilot-wave theory.
\newblock {\em Journal of Physics A: Mathematical and Theoretical},
  47(39):395306, sep 2014.
\newblock \href {https://doi.org/10.1088/1751-8113/47/39/395306}
  {\path{doi:10.1088/1751-8113/47/39/395306}}.

\bibitem{lustosa2021quantum}
Francisco~Bento Lustosa, Samuel Colin, and Santiago~E Perez~Bergliaffa.
\newblock Quantum relaxation in a system of harmonic oscillators with
  time-dependent coupling.
\newblock {\em Proceedings of the Royal Society A: Mathematical, Physical and
  Engineering Sciences}, 477(2248), 2021.
\newblock \href {https://doi.org/10.1098/rspa.2020.0606}
  {\path{doi:10.1098/rspa.2020.0606}}.

\bibitem{lustosa2023evolution}
Francisco~Bento Lustosa, Nelson Pinto-Neto, and Antony Valentini.
\newblock Evolution of quantum non-equilibrium for coupled harmonic
  oscillators.
\newblock {\em Proceedings of the Royal Society A: Mathematical, Physical and
  Engineering Sciences}, 479(2269), 2023.
\newblock \href {https://doi.org/10.1098/rspa.2022.0411}
  {\path{doi:10.1098/rspa.2022.0411}}.

\bibitem{valentini2010inflationary}
Antony Valentini.
\newblock Inflationary cosmology as a probe of primordial quantum mechanics.
\newblock {\em Phys. Rev. D}, 82:063513, Sep 2010.
\newblock URL: \url{https://link.aps.org/doi/10.1103/PhysRevD.82.063513}, \href
  {https://doi.org/10.1103/PhysRevD.82.063513}
  {\path{doi:10.1103/PhysRevD.82.063513}}.

\bibitem{colin2013mechanism}
Samuel Colin and Antony Valentini.
\newblock Mechanism for the suppression of quantum noise at large scales on
  expanding space.
\newblock {\em Phys. Rev. D}, 88:103515, Nov 2013.
\newblock URL: \url{https://link.aps.org/doi/10.1103/PhysRevD.88.103515}, \href
  {https://doi.org/10.1103/PhysRevD.88.103515}
  {\path{doi:10.1103/PhysRevD.88.103515}}.

\bibitem{colin2015primordial}
Samuel Colin and Antony Valentini.
\newblock Primordial quantum nonequilibrium and large-scale cosmic anomalies.
\newblock {\em Phys. Rev. D}, 92:043520, Aug 2015.
\newblock URL: \url{https://link.aps.org/doi/10.1103/PhysRevD.92.043520}, \href
  {https://doi.org/10.1103/PhysRevD.92.043520}
  {\path{doi:10.1103/PhysRevD.92.043520}}.

\bibitem{colin2016robust}
Samuel Colin and Antony Valentini.
\newblock Robust predictions for the large-scale cosmological power deficit
  from primordial quantum nonequilibrium.
\newblock {\em International Journal of Modern Physics D}, 25(06):1650068,
  2016.
\newblock URL: \url{https://doi.org/10.1142/S0218271816500681}.

\bibitem{underwood2015quantum}
Nicolas~G. Underwood and Antony Valentini.
\newblock Quantum field theory of relic nonequilibrium systems.
\newblock {\em Phys. Rev. D}, 92:063531, Sep 2015.
\newblock URL: \url{https://link.aps.org/doi/10.1103/PhysRevD.92.063531}, \href
  {https://doi.org/10.1103/PhysRevD.92.063531}
  {\path{doi:10.1103/PhysRevD.92.063531}}.

\bibitem{valentini2025toward}
Antony Valentini and Mira Varma.
\newblock Toward a test of the born rule in high-energy collisions.
\newblock {\em Phys. Rev. D}, 112:112024, Dec 2025.
\newblock URL: \url{https://link.aps.org/doi/10.1103/4llk-pt1j}, \href
  {https://doi.org/10.1103/4llk-pt1j} {\path{doi:10.1103/4llk-pt1j}}.

\bibitem{kandhadai2020mechanism}
Adithya Kandhadai and Antony Valentini.
\newblock Mechanism for nonlocal information flow from black holes.
\newblock {\em International Journal of Modern Physics A}, 35(06):2050031,
  2020.
\newblock \href {https://doi.org/10.1142/S0217751X20500311}
  {\path{doi:10.1142/S0217751X20500311}}.

\bibitem{valentini2023beyond}
Antony Valentini.
\newblock {Beyond the Born Rule in Quantum Gravity}.
\newblock {\em Found. Phys.}, 53(1):6, 2023.
\newblock \href {https://arxiv.org/abs/2212.12175} {\path{arXiv:2212.12175}},
  \href {https://doi.org/10.1007/s10701-022-00635-0}
  {\path{doi:10.1007/s10701-022-00635-0}}.

\bibitem{WheelerQG}
J.~A. Wheeler.
\newblock Relativity groups and topology.
\newblock {\em In Relativity groups and Topology, 1963 Les Holches Lectures},
  1970.

\bibitem{colistete1998singularities}
Roberto Colistete, J\'ulio~C. Fabris, and Nelson Pinto-Neto.
\newblock Singularities and the classical limit in quantum cosmology with
  scalar fields.
\newblock {\em Phys. Rev. D}, 57:4707--4717, Apr 1998.
\newblock URL: \url{https://link.aps.org/doi/10.1103/PhysRevD.57.4707}, \href
  {https://doi.org/10.1103/PhysRevD.57.4707}
  {\path{doi:10.1103/PhysRevD.57.4707}}.

\bibitem{bohm1954model}
D.~Bohm and J.~P. Vigier.
\newblock Model of the causal interpretation of quantum theory in terms of a
  fluid with irregular fluctuations.
\newblock {\em Phys. Rev.}, 96:208--216, Oct 1954.
\newblock URL: \url{https://link.aps.org/doi/10.1103/PhysRev.96.208}, \href
  {https://doi.org/10.1103/PhysRev.96.208} {\path{doi:10.1103/PhysRev.96.208}}.

\end{thebibliography}

\end{document}